**Software paper for submission to the Journal of Open Research Software**

**(1) Overview**

**Title**

# Wrangler for the Emergency Events Database: A Tool for Geocoding and Analysis of a Global Disaster Dataset

**Paper Authors**
1. Kripa, Ram M. ;
2. Ramesh, Nandini ;
3. Boos, William R.

**Paper Author Roles and Affiliations**
1. Ram M. Kripa: Department of Earth and Planetary Science, University of California, Berkeley
2. Dr. Nandini Ramesh: ARC Centre for Data Analytics for Resources and the Environment, The University of Sydney
3. Prof. William R. Boos: Department of Earth and Planetary Science, University of California, Berkeley; Climate and Ecosystem Sciences Division, Lawrence Berkeley National Laboratory

**Abstract**
There is an increasing need for precise location information on historical disasters, such as mass casualty events caused by weather or earthquakes, but existing disaster datasets often do not provide geographic coordinates of past events.  Here we describe a new tool, the Wrangler for the Emergency Events Database (WEED), that associates latitude and longitude coordinates with entries in the widely used Emergency Events Database (EM-DAT). WEED takes as input records from EM-DAT, and geocodes the list of cities, states, and other location types associated with a given disaster using the R language with the GeoNames web service.  Error processing is performed, and users are given the ability to customize the logic used in geocoding; the open-source nature of the tool also allows more general customization or extension by users. This tool provides researchers the ability to easily prepare EM-DAT data for analysis with geophysical, hydrological, and other geospatial variables.

**Introduction**

With the large number of people affected by natural disasters every year, there is a need for high-quality location information on current and historical disasters. Such location information is crucial for planning disaster response [1] and for understanding the relationship between environmental conditions and human casualties [2].

One of the most widely used sources of disaster data is the Emergency Events Database (EM-DAT) [3], which contains geographic, temporal, human, and economic data pertaining to more than 20,000 mass disasters from 1900 to the present (e.g., Table 0). EM-DAT was created and is maintained by the Centre for Research on the Epidemiology of Disasters (CRED) at the Université catholique de Louvain, to help mitigate the impact of disasters on vulnerable populations through systematic collection and analysis of data. For each disaster, the database may contain estimates of human impact (number of people killed and injured), economic impact (estimated monetary damage), and aid contributions. It also contains fields for the time and location of each disaster, but the location most often consists of plain text strings of locations, rendering spatial analysis difficult. Latitude and longitude fields exist in the database, but are sparsely populated and, when complete, associate only one spatial point with many disasters that affect multiple places.

Two previous efforts have been made to remedy this deficiency in geographic location data in EM-DAT. The first of these, the Geocoded Disasters (GDIS) dataset, provides polygons associated with 9,924 disasters occurring between 1960 and 2018 [4]. The GDIS dataset relied on manual coding, taking over five years to complete, and hence will not automatically include new data from 2019 onward. Additionally, its data format of location polygons may be more computationally intensive and involved to process by subsequent analyses than simple latitude-longitude coordinates. Furthermore, since the polygons in GDIS were obtained by associating EM-DAT location names with boundaries from the Global Administrative Areas [GADM] database, they do not represent a precisely determined disaster boundary. The second effort to better quantify disaster data location is included with newer entries in EM-DAT. Recently, the maintainers of EM-DAT began to include codes in the dataset that refer to affected administrative units in GAUL (Global Administrative Unit Layers). Work on providing user access to the specific shapefiles is ongoing. Similar to the GDIS effort, this functionality is limited by its temporal coverage, as this data only exists for disasters that occured in the year 2000 or later; it also requires processing with Geographic Information System (GIS) software.

Here we introduce a new method to geocode disaster information from EM-DAT. The Wrangler for the Emergency Events Database (WEED) is an R package, available on the Comprehensive R Archive Network (CRAN), that retrieves latitude-longitude coordinates of disasters from EM-DAT on demand. WEED is meant to be complementary to GDIS and the native shapefiles provided in EM-DAT for some recent disasters, with WEED designed to maximize coverage and dynamism while minimizing the need for specialty GIS software and computing power to analyze polygons. Hence, it dynamically queries the GeoNames API to retrieve coordinates for each location. It outputs latitude-longitude coordinates for each location name affected by a disaster, allowing multiple points to be obtained when more than one geographic location is affected by a disaster. This paper describes WEED, facilitating its use in geocoding any set of disasters in past, present, and future versions of EM-DAT.

**Implementation and architecture**

The structure of WEED is best explained through a sample workflow (with package functions italicized):

1. Downloading native EM-DAT data and loading it into WEED using *read_emdat*
2. Changing the unit of analysis from disasters to individual locations affected by the disasters with *split_locations*
3. Characterizing output using *percent_located_locations* and *percent_located_disasters*
4. Geocoding using *geocode* or *geocode_batches*
5. Characterization of the resulting geocoded data using *percent_located_locations* and *percent_located_disasters*
6. Conducting spatial analysis, e.g. using *located_in_box* or *located_in_shapefile*

1. **Loading data**

First, one downloads the subset of EM-DAT required for one's analysis of choice using the public EM-DAT query tool, which returns data in Microsoft Excel format. The read_emdat function then allows the dataset to be loaded with or without its metadata. The metadata, included as a header in the Excel spreadsheet output by EM-DAT, contains information regarding the EM-DAT query, including its timestamp, version, and the type of request; this metadata can be useful for documentation and reproducibility.  A sample function call is:

```
sample_data <- read_emdat(here("data",
"emdat_public_2021_01_12_full.xlsx"))
sample_df <- sample_data[["disaster_data"]]
```

Some of the most commonly used data fields from a sample EM-DAT subset can be seen in Table 1.

At this point one usually conducts a brief examination of the loaded data. The fact that only a small fraction of disasters have existing quantitative location data, in the form of latitude and longitude coordinates (Table 1 in blue) or shapefiles, hampers geospatial analysis and motivated development of the WEED package.

## 2. Changing the unit of analysis

An issue that must be addressed in the geocoding process is the presence of multiple locations per disaster (e.g., Table 1 in orange). The recommended WEED workflow is to change the unit of analysis from "one row per disaster" to "one row per disaster-location pair", with the disaster number persisted in each row serving to link locations that all correspond to the same disaster.

Changing the unit of analysis in this way is complicated by the fact that EM-DAT contains only one location field for each disaster record, so potentially many locations associated with each disaster are contained in a single text string. Furthermore, the locations in this string are represented in a wide variety of ways in EM-DAT, and the delimiter used to separate individual locations varies across records. All of this is presumably due to the fact that EM-DAT is a database that has been manually filled row-by-row with no consistent formatting standard for location names. For example, U.S. states are sometimes abbreviated and sometimes spelled out, and city names are only sometimes accompanied by the names of their states. Methods used in EM-DAT to represent a list of locations (here called A, B and C) include "A, B, and C", "(1) A (2) B (3) C", and "A (B and C)".

**Sample Data**

| Dis.No | Country | Disaster.Type | Location | Latitude | Longitude | Total.Deaths | CPI |
|---|---|---|---|---|---|---|---|
| 2000-0919-USA | United States of America (the) | Storm | Alabama, Georgia, Louisiana, North Carolina, South Carolina, Tennessee, Virginia, New York, Pennsylvania, Massachussetts provinces | | | 4 | 67.355759 |
| 1928-0024-CAN | Canada | Earthquake | Burin Peninsula, Newfoundland | 48.60 N | 58.00 W | 27 | 6.731507 |
| 1998-0212-USA | United States of America (the) | Extreme temperature | Arizona, Florida, Colorado, Texas, Oklahoma, LA | | | 130 | 63.760455 |

Table 1: A sample dataframe of 3 records drawn from the EMDAT dataset, with columns of interest highlighted. The coverage of the dataset in terms of text locations is quite comprehensive, but only a few disasters have entries in the Latitude and Longitude columns.

The process of changing the unit of analysis from disasters to disaster-location pairs will henceforth be called "locationizing" and is illustrated in Table 2 for the first disaster taken from Table 1. Locationizing is performed using the `split_locations` function:

```
locationized_df <- sample_df %>%
  split_locations(column_name = "Location")
```

| Locationized Sample Data | | | | | | |
|---|---|---|---|---|---|---|
| Dis No | Country | Disaster Type | Latitude | Longitude | location_word | uncertain_location_specificity |
| 2000-0919-USA | United States of America (the) | Storm | | | alabama | FALSE |
| 2000-0919-USA | United States of America (the) | Storm | | | georgia | FALSE |
| 2000-0919-USA | United States of America (the) | Storm | | | louisiana | FALSE |
| 2000-0919-USA | United States of America (the) | Storm | | | north carolina | FALSE |
| 2000-0919-USA | United States of America (the) | Storm | | | south carolina | FALSE |
| 2000-0919-USA | United States of America (the) | Storm | | | tennessee | FALSE |

Table 2 : Subset of the locationized sample dataframe. Every row now corresponds to a unique pair of one disaster (Dis No) and one location (location_word), rather than one disaster with at least one location.

The `split_locations` function uses a method of splitting that incorporates a default set of delimiters (e.g. commas) and generic names for geographic administrative levels (e.g. "state" and "province"), with the user allowed to specify additional ones in the optional parameters `joiner_regex` and `dummy_words`, respectively. In the example function call displayed above, we use the default parameters and so do not specify either of these.

For example, one location string in our dataset (Table 1) includes the string "New York, Pennsylvania, and Massachusetts provinces" (the word "provinces" is used even though these are U.S. states). Dummy words included in the package by default include "state(s)", "province(s)", "town(s)", and other words that specify the administrative level. The word "provinces" is removed as one of the default dummy words, and the commas and the word "and" are removed as default delimiters. The resulting split string yields three `location_words`, "new york", "pennsylvania", and "massachusetts".

When parentheses are present in location strings, like "Berkeley (California)" or "California (Berkeley, Emeryville, Alameda)", they may indicate differing levels of specificity. We therefore cannot infer the specificity of each of these locations just by their position relative to their associated parentheses. Hence, we add an uncertainty flag called `uncertain_location_specificity` (Table 2 in green).

## 3. Characterization of existing location data

The locationized data (contained in an R data frame) is compatible with some brief exploratory package functions that allow for easy visualization of the fraction of the dataset that already contains geographic coordinates.

The function percent_located_locations displays the proportion of locations (i.e. location-disaster pairs) that contain latitude-longitude coordinates. Because no geocoding has yet been performed using WEED, about 88 percent (16 out of 18) of location-disaster pairs in this sample dataset have no associated latitude and longitude data (Figure 1). A similar function exists for disasters (which can have many locations), showing that 67 percent of disasters in this sample dataset have no latitude-longitude coordinates (2 out of 3).

```
locationized_df %>%
  percent_located_locations(lat_column = "Latitude",
                            lng_column = "Longitude")
locationized_df %>%
  percent_located_disasters(lat_column = "Latitude",
                            lng_column = "Longitude")
```

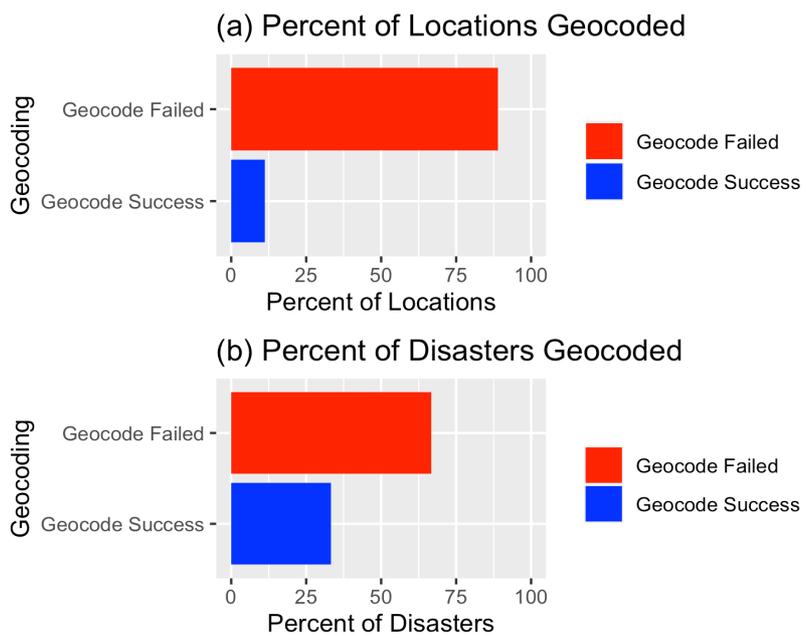

Figure 1: Bar plots of the percentage of a) location-disaster pairs (left) and b) disasters (right) for which the latitude and longitude columns are empty.

## 4. Geocoding

WEED uses the Geonames API to geocode each location. To use this functionality, one must first create a free account on Geonames and then supply a username to the geocode function in WEED. This function comes with a few options, depending on the kind of analysis that is being performed. Users can employ the *n_results* parameter to get the *n* closest matches to the input location and decide which one to use. The *unwrap* parameter allows users to persist the geocoded data in a nested data frame structure, which may be useful when exporting for further analysis, or to obtain the geocoded data in unwrapped form, with each latitude and longitude assigned in separate fields (lat1, lng1, lat2, lng2, etc.)

```
geocoded_df <- locationized_df %>%
  geocode(unwrap = FALSE, geonames_username = sample_username)
```

As seen in Table 3, the geographic coordinates generated by geocoding (in green) are considerably more complete than the coordinates provided in the original dataset (in blue).

| Geocoded Data | | | | | | |
|---|---|---|---|---|---|---|
| Dis No | Country | Latitude | Longitude | location_word | lat | lng |
| 2000-0919-USA | United States of America (the) | | | alabama | 34.60739 | -86.97977 |
| 2000-0919-USA | United States of America (the) | | | georgia | 33.69277 | -84.39957 |
| 2000-0919-USA | United States of America (the) | | | louisiana | 30.12595 | -92.00939 |
| 2000-0919-USA | United States of America (the) | | | north carolina | 34.00071 | -81.03481 |
| 2000-0919-USA | United States of America (the) | | | south carolina | 34.00071 | -81.03481 |
| 2000-0919-USA | United States of America (the) | | | tennessee | 35.80000 | -86.50000 |

Table 3 : Subset of the sample dataframe after geocoding. Note the superior coverage of our new columns, lat and long, over the latitude and longitude columns provided by EMDAT.

If the input dataset has more than 1,000 records (rows) after locationizing it is advisable to use the function *geocode_batches* which includes a latency period between batches to comply with Geonames query rate limits. In addition to geocode parameters, *geocode_batches* takes the additional parameters *batch_size* (number of records) and *wait_time* (in seconds). Geonames currently allows 1,000 queries per hour and a maximum of 20,000 queries per day under its free access plan. Hence, the batching default parameters (*batch_size* and *wait_time*) have been set to avoid exceeding this threshold.

```
geocoded_df2 <- locationized_df %>%
  geocode_batches(batch_size = 990, wait_time = 4800, unwrap =
FALSE, geonames_username = sample_username)
```

## 5. Characterization of geocoding results

Now we repeat use of the percent_located_locations and percent_located_disasters functions to visualize the success of the geocoding.

```
geocoded_df %>%
  percent_located_locations()
geocoded_df %>%
  percent_located_disasters(how = "any")
```

This shows that 90 percent of the location-disaster pairs in the sample dataset have been successfully geocoded (Figure 2a).

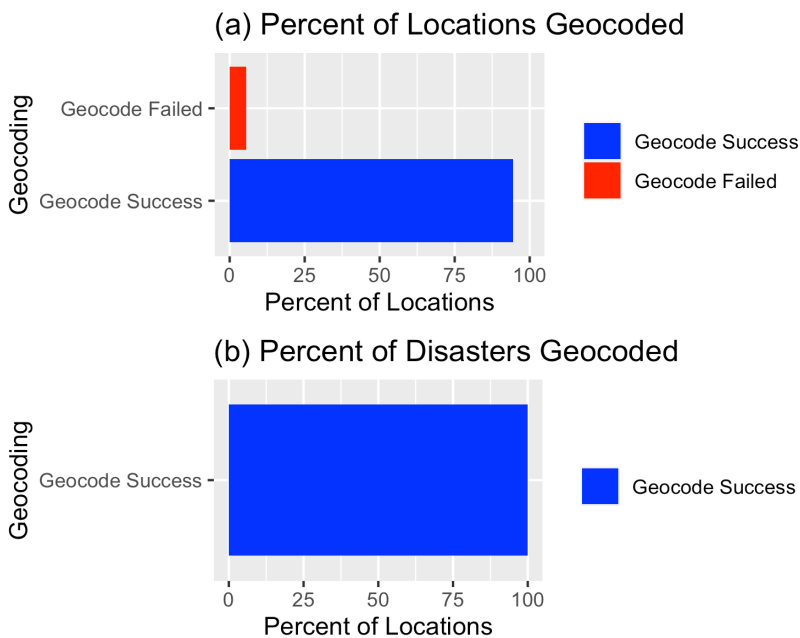

Figure 2 : The percentage of a) location-disaster pairings for which the latitude and longitude fields are filled in the geocoded dataset, and b) disasters for which at least one of its constituent locations has the latitude and longitude fields filled.

When determining the percentage of disasters that have been located, we now must decide whether we will count a disaster as "located" if at least one of its constituent locations has valid geographic coordinates, if all of those locations have coordinates, or some other number. This choice did not need to be made when we examined the percent located prior to geocoding because the native EM-DAT records contain only one latitude-longitude coordinate for each disaster, even if many place names are in the location string. The criterion used for determining whether a disaster was located are specified in the *how* parameter of the `percent_located_disasters` function; using the value "any" will count a disaster as located if at least one of its constituent locations has a valid geographic coordinate, while using "all" requires every constituent location to be geocoded.

For our sample dataset, we see that 100% of disasters have at least one location that was successfully geocode (Fig. 3b). User-defined aggregation functions may also be passed to the *how* parameter. For example, if we define a function that returns true if 50% or more of the locations corresponding to a disaster had been geocoded, then that function could be passed as the *how* parameter.

## 6. Elementary spatial analysis

WEED includes a few simple functions that allow for elementary spatial analysis of the geocoded data, motivated by the fact that a common task in analysis workflows is determining whether a geographic point lies within some defined region.

If the defined region of interest is a rectilinear box on a latitude-longitude grid, one can use the `located_in_box` function. For example, one can define a box defined using these two points as the top-left and bottom-right corners:

```
tllat = 40 # top left latitude
tllng = -119 # top left longitude
brlat = 35 # bottom right latitude
brlng = -75 # bottom right longitude
```

To determine whether each of the disaster-location pairs in a dataset falls within this bounding box, WEED provides the `located_in_box` function. This function creates a new column called 'in_box' that stores a Boolean value for each record, and the resulting output of this function (6th Column of Table 4) can be easily piped with elementary visual analysis on a map (Figure 3).

```
inbox_df <- geocoded_df %>%
  located_in_box(top_left_lat = tllat, top_left_lng = tllng,
bottom_right_lat = brlat, bottom_right_lng = brlng)
```

**In Box and In Shape Data**

| Dis No | Country | location_word | lat | lng | in_box | in_shape |
|---|---|---|---|---|---|---|
| 2000-0919-USA | United States of America (the) | alabama | 34.60739 | -86.97977 | FALSE | FALSE |
| 2000-0919-USA | United States of America (the) | georgia | 33.69277 | -84.39957 | FALSE | FALSE |
| 2000-0919-USA | United States of America (the) | louisiana | 30.12595 | -92.00939 | FALSE | FALSE |
| 2000-0919-USA | United States of America (the) | north carolina | 34.00071 | -81.03481 | FALSE | FALSE |
| 2000-0919-USA | United States of America (the) | south carolina | 34.00071 | -81.03481 | FALSE | FALSE |
| 2000-0919-USA | United States of America (the) | tennessee | 35.80000 | -86.50000 | TRUE | FALSE |

Table 4 : Subset of the sample dataframe after applying the located_in_box and located_in_shapefile methods. Note the new columns in_box and in_shape that indicate whether the given location is within the lat-long box and polygon respectively.

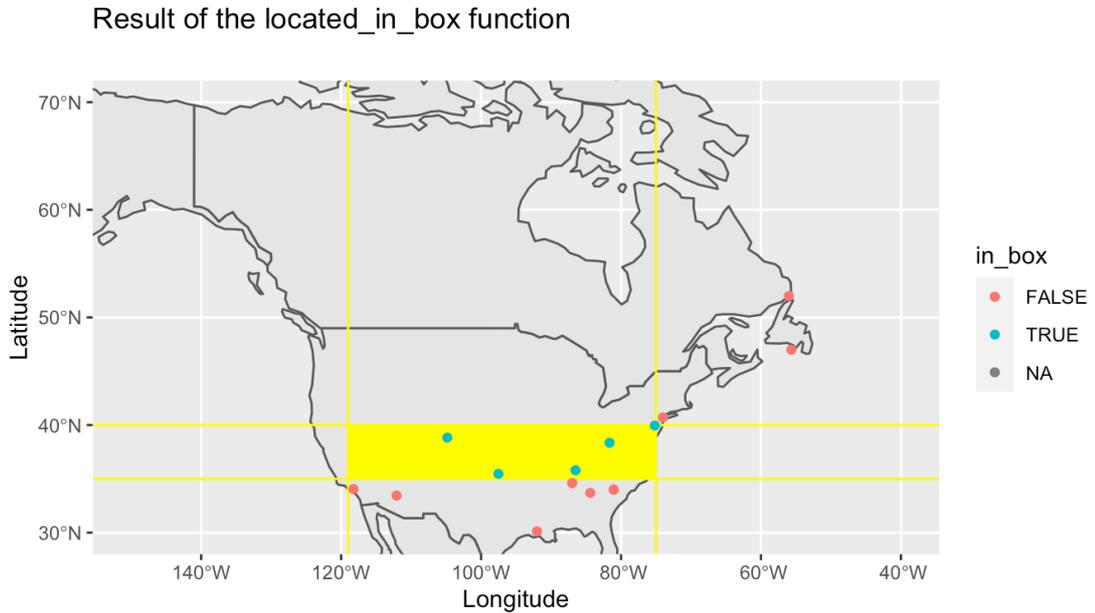

Figure 3 : Map with all of the sample data points plotted, as well as the bounding box used above.

If the region of interest is a polygon specified as a shapefile, one can instead use the `located_in_shapefile` function. For example, if the shapefile contains the boundaries of the state of California. The located_in_shapefile function can take in polygons through the shapefile parameter, or even extract the polygons from a shapefile whose name is passed as the shapefile_name parameter. As with located_in_box, it appends a new column, 'in_shape' to the dataset (7th Column of Table 4), allowing for easy visualization (Figure 4).

```
in_shape_df <- inbox_df %>%
  located_in_shapefile(shapefile = map_california)
```

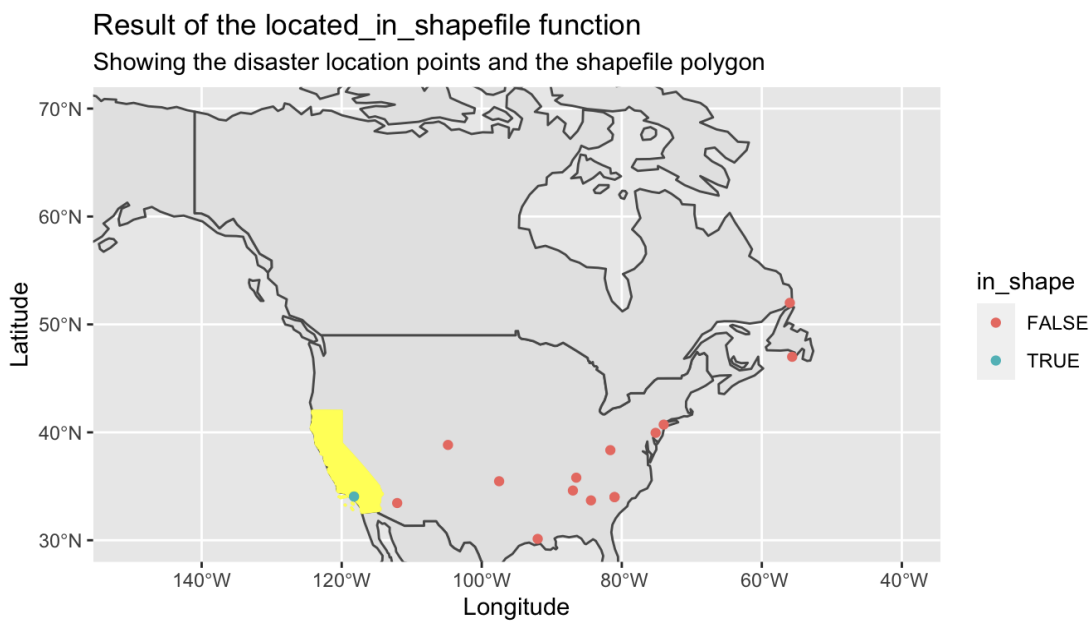

Figure 4 : Map with all of the sample data points plotted, as well as the shapefile polygon used. Points identified as being inside the shapefile are shown in blue.

These functions are intended to aid in some common geocoding use cases and serve as a starting point for further analyses.

**Quality control**
The WEED package has been tested by CRAN as part of its submission process, as well as by the developer with unit tests. Relevant examples specific to each function can be found in the package documentation, also available on CRAN, and in the R Console.

**(2) Availability**

*Operating system*
Compatible with MacOS (on both x86 and ARM architectures), Windows, Linux (Fedora, Debian), and Solaris, as tested by CRAN.

*Programming language*
Compatible with R versions >= 4.0.0

*Additional system requirements*
The user requires a geonames account and internet connection to utilize geocoding features of the package.

*Dependencies*
The following R packages are required: readxl, dplyr, magrittr, tidytext, stringr, tibble, geonames, countrycode, purrr, tidyr, forcats, ggplot2, rgeos, sf, here

*List of contributors*
Ram M. Kripa: Developer

*Software location:*
    **Archive**
        *Name:* CRAN
        *Persistent identifier:* https://CRAN.R-project.org/package=weed
        *Licence:* https://cran.r-project.org/web/packages/weed/LICENSE
        *Publisher:* Ram M. Kripa
        *Version published:* 1.1.1
        *Date published:* 07/07/2021
    **Code repository**
        *Name:* Github
        *Identifier:* https://github.com/rammkripa/weed
        *Licence:* https://github.com/rammkripa/weed/blob/master/LICENSE
        *Date published:* 07/06/2021

*Language*
R

## (3) Reuse potential

The WEED package can be used for analysis of the spatio-temporal patterns of disasters by efficiently associating geographic coordinates with disasters obtained from the widely used EM-DAT dataset. The approach taken by WEED is distinct from the manual method used by one previous effort [4], which resulted in a static dataset limited to the years 1960-2018; it also differs from the in-progress geocoding being provided by the native EM-DAT, which extends back in time to only the year 2000. WEED further contrasts with those two approaches by providing sets of latitude-longitude coordinates for each disaster, rather than polygons or shapefiles; for some purposes this may be less accurate, but it also allows for easier processing outside of GIS software. By providing dynamic, on-demand geocoding with easily customizable treatment of local place names, this tool may accelerate progress in understanding and eventually preventing disasters.

## Acknowledgements

--

## Funding statement

--

## Competing interests

The authors declare that they have no competing interests.

---